\documentclass[journal]{IEEEtran}
\usepackage{gensymb}
\usepackage{amsmath}
\usepackage{comment}
\usepackage{graphicx}
\usepackage{threeparttable}
\usepackage{tabularx, mathtools}
\usepackage{url}
\usepackage{breakurl}

\usepackage{enumerate}
\usepackage{hyperref}
\hypersetup{
    colorlinks=true,
    linkcolor=blue,
    urlcolor=blue,
    citecolor=blue,
    filecolor=blue
    }
\ifCLASSINFOpdf
\else
\fi

\begin{document}
\bstctlcite{IEEEexample:BSTcontrol}

%
\title{
Navigation Method Enhancing Music Listening Experience by Stimulating Both Neck Sides with Modulated Musical Vibration
}

%
%
%

\author
{
Yusuke~Yamazaki 
and~Shoichi~Hasegawa.
\thanks{
Y. Yamazaki and S. Hasegawa are with the Department of Information and Communications Engineering, School of Engineering, Tokyo Institute of Technology, Japan,
e-mail: \{yus988, hase\}@haselab.net}

}


%
%


\markboth{}%
{Shell \MakeLowercase{\textit{et al.}}: Navigation Method Enhancing Music Listening Experience by Stimulating Both Neck Sides with Modulated Music Vibration}
%



\maketitle




%

\begin{abstract}
We propose a method that stimulates musical vibration (generated from and synchronized with musical signals), modulated by the direction and distance to the target, on both sides of a user's neck with Hapbeat, a necklace-type haptic device.
We conducted three experiments to confirm that the proposed method can achieve both haptic navigation and enhance the music-listening experience.
Experiment 1 consisted of conducting a questionnaire survey to examine the effect of stimulating musical vibrations.
Experiment 2 evaluated the accuracy (deg) of users' ability to adjust their direction toward a target using the proposed method.
Experiment 3 examined the ability of four different navigation methods by performing navigation tasks in a virtual environment.
The results of the experiments showed that stimulating musical vibration enhanced the music-listening experience, and that the proposed method is able to provide sufficient information to guide the users: accuracy in identifying directions was about 20\textdegree, participants reached the target in all navigation tasks, and in about 80\% of all trials participants reached the target using the shortest route.
Furthermore, the proposed method succeeded in conveying distance information, and Hapbeat can be combined with conventional navigation methods without interfering with music listening.
\end{abstract}
\IEEEpeerreviewmaketitle

\section*{Copyright Notice} 
DOI: \href{https://ieeexplore.ieee.org/document/10098889}{10.1109/TOH.2023.3266194}
\copyright2023 IEEE. Personal use of this material is permitted. Permission from IEEE must be obtained for all other uses, in any current or future media, including reprinting/republishing this material for advertising or promotional purposes, creating new collective works, for resale or redistribution to servers or lists, or reuse of any copyrighted component of this work in other works.

\section{Introduction}
\label{sec_intro}
Listening to music while walking is a daily activity for many people \cite{haas2018can}.
The widespread use of smartphones and wireless headphones with active noise cancellation and transparency mode has made it easier for pedestrians to enjoy high-quality music with their favorite listening conditions.
Pedestrians also tend to use smartphone navigation applications when visiting new places.
Many such applications provide voice guidance so that users can reach their destination without looking at the screen while walking.
However, such voice guidance can disturb music listening.
With this background, researchers have proposed navigation methods that indicate the direction to be travelled by modulating music, so as not to disturb music listening while navigating \cite{jones2008ontrack, yamano2012eyesound, heller2018navigatone, heller2020attracktion}.
These studies have shown that localization using music modulation can provide the same level of navigation capability as voice guidance.
\par
However, the studies have not discussed improving the music-listening experience.
Listening to music is an experience that involves feeling music with the entire body, not just the auditory senses \cite{reybrouck2019music}.
Researchers have reported that sound energy emitted from live music and loudspeakers can vibrate the human body through the air, as well as structures such as chairs and floors \cite{merchel2013music, takahashi2002some}.
According to recent studies \cite{merchel2018auditory, hove2020feel}, the feeling on the body of vibrations generated from and synchronized with musical signals (hereafter, such vibration is referred to as musical vibrations), especially low-frequency vibrations, plays an important role in the music-listening experience.
To experience musical vibrations, we first proposed a vibration generation mechanism that can transmit powerful low-frequency vibrations over a wide area with a small device using motors and a thread \cite{yamazaki2016tension}.
Then we implemented a necklace-type haptic device, Hapbeat, using our proposed mechanism that is easy to use when walking \cite{yamazaki2022implementation}.
While Hapbeat can enhance the music-listening experience by transmitting low-frequency vibrations over a wide area, it can also deliver localization information by independently controlling the left and right sides of the ribbon.
For example, we proposed a method for presenting a target's location by vibrating both sides of the neck with a stereo sine wave where the stereo balance (i.e., the amplitude ratio of the left-right channel) and total amplitude are modulated by the direction and distance toward the target \cite{yamazaki2019neck}.
\par
With this background, we conceived that stimulating musical vibration modulated by the direction and distance to the destination with Hapbeat can achieve both haptic navigation and enhance the music-listening experience.
We made the following hypotheses and tested them in this paper.
\begin{itemize}
\item
H1: Modulating the stereo balance of musical vibrations can convey enough information for navigation.
\item
H2: Stimulating musical vibration while walking enhances the music-listening experience.
\item
H3: Stimulating musical vibration while walking does not interfere with the previously proposed navigation method that modulates musical sound.
\item
H4: Navigating by modulating musical vibration is preferred as a music-listening experience over modulating music sound.
\item
H5: Presenting distance information by modulating the amplitude of the musical vibrations makes it easier for users to understand the navigation.
\item
H6: Presenting distance information by modulating the amplitude of the musical vibrations negatively impacts the music-listening experience.
\end{itemize}
\section{Related work}
\subsection{Non-visual navigation methods}
Since looking at a screen while walking poses various safety issues, researchers have proposed many navigation methods that use audio and haptic feedback instead of vision.
\subsubsection{Navigation by audio (music) modulation}
\label{sec_rwMzNavi}
The most popular audio navigation method uses voice guidance through verbal speech, typified by Google Maps.
The method does not require training and can provide detailed information, but is not suitable when the user is performing a cognitively demanding task because of the language processing required \cite{klatzky2006cognitive}.
Holland \textit{et al.} \cite{holland2002audiogps} proposed AudioGPS, a voice navigation system with low cognitive load.
AudioGPS modulates the stereo balance of short instrumental sounds to present the direction and the playing interval to present the distance as navigation cues.
Their study showed that AudioGPS successfully and intuitively guides users to their destinations.
\par
Since then, researchers have proposed various navigation methods that modulate music to make auditory navigation more natural and usable in daily life.
Strachan \textit{et al.} \cite{strachan2005gpstunes} proposed GpsTune, a beacon guidance system that provides the direction to the destination by modulating the stereo balance of the music and the distance by modulating the volume. Jones \textit{et al.} \cite{jones2008ontrack} created a similar system, ONTRACK, and conducted a detailed user evaluation.
They reported that 90\% of ONTRACK users reached their destination with ONTRACK-only navigation when directions were presented at a resolution of 30$^\circ$ to the destination.
In contrast, they reported that the presentation of distance by volume modulation did not work well; continuous changes in volume were difficult to understand, and the structure of the music and the fade-out at the end of the piece caused the volume to change independently of the distance.
Yamano \textit{et al.} \cite{yamano2012eyesound} proposed EyeSound, which modulates the phase difference between the music's left and right sound to change the sound localization and provide a directional presentation.
EyeSound users could correctly perceive the direction over 80\% of the correct rate with a resolution of 45\textdegree~(navigation applications were not evaluated).
\par
These studies show that music modulation can be used for navigational purposes, but they do not discuss whether the music-listening experience is natural.
In particular, modulating the stereo balance of the music inevitably results in a larger volume difference between the left and right sides when the target is right beside you, affecting perception of the music \cite{kimura1964left}.
To facilitate navigation using a more natural music-listening experience, Heller \textit{et al.} \cite{heller2018navigatone, heller2020attracktion} proposed the method NavigaTone, which only modulated specific music tracks, such as vocals and drums.
They initially evaluated the NavigaTone method by asking participants to identify sound sources randomly placed at resolutions of 15\textdegree~in front of them and 45\textdegree~behind them.
According to the results, when the authors counted off-by-one answers as correct (i.e., within a 30\textdegree~error margin), NavigaTone's success rate was 86\% while that of the conventional stereo panning method was 90\%,
Based on participant interviews, Heller \textit{et al.} also reported that almost all participants enjoyed listening to music more with NavigaTone than with stereo panning.
In a subsequent study, \cite{heller2020attracktion} conducted navigation experiments in real-world walking in order to evaluate the potential of their proposed method compared to conventional turn-by-turn instructions.
The result showed that the two methods performed comparably in terms of path efficiency, navigation errors, and mental workload.
However, it should be noted that the method does not specify distance information.
\subsubsection{Navigation by haptic feedback}
\label{rw_hapticNavi}
Many studies have reported using haptic feedback for navigation and direction presentation.
A common approach attaches haptic devices with arrays of eccentric motors or linear vibrators to the user's body.
Direction and distance information is delivered by controlling the vibrating positions and vibration patterns.
Various researchers have adapted this approach for different parts of the body \cite{bosman2003gentleguide,salzer2010vibrotactor,erp2005waypoint,schaack2019haptic}.
The advantage of the method is that the spatial distribution of the vibratory stimuli allows for the intuitive conveyance of two-dimensional information.
In addition, there are methods using a single vibrator that turn the vibration on and off only when the user is facing the desired direction \cite{marston2007nonvisual} and methods that combine sound and vibration patterns to present distance and direction \cite{fujimoto2014non}.
The advantage of these methods is that they can use ordinary smartphones and do not require special equipment.
In addition to the methods using vibrators, other methods have been proposed using thermal perception \cite{di2019haptic}, skin stretch force \cite{chinello2017design}, changing a device's shape \cite{spiers2016design}, and pseudo-attraction force \cite{amemiya2009haptic}, all of which intuitively present directional information.
To our knowledge, however, no research has yet been conducted that focuses primarily on achieving both haptic navigation and enhancing the music-listening experience.
\subsection{Enhancing the music-listening experience through musical vibrations}
Several studies have shown that musical vibration stimulation in the low-frequency range positively influences the music-listening experience.
Merchel \textit{et al.} \cite{merchel2018auditory} had subjects sit in a whole-body vibration device and evaluated the quality of their experience when listening to music with and without vibration based on audio signals.
Four genres of music were included in their experiment, and subjects preferred the listening experience with vibration compared to that without vibration for all music, in particular giving a high rating for pop music with strong frequency components below 20 Hz.
Hove \textit{et al.} \cite{hove2020feel} hypothesized that the simultaneous stimulation of sound and low-frequency vibrations would enhance subjects’ groove to the music and tested the hypothesis using SUBPAC M2X (www.subpac.com), a backpack-type haptic device with a large built-in linear vibrator.
The subjects were instructed to tap their fingers to the music and asked about their impressions of music listening.
The results showed that with musical vibration, the intensity of the subjects’ tapping became stronger and their groove and enjoyment of the music improved, indicating the effect of low-frequency vibration.
However, the seated situation might have enabled participants in these studies to concentrate more on music listening, which differs from the effect of the walking situation in the sense that movement and stimulation from the feet when walking could distract participants from music listening.
\par
In relation to groove, Senn \textit{et al.} \cite{senn2020experience} wrote ``the definition of groove as music listeners' inner urge to move their bodies in response to the music is unanimously accepted in music psychology, but a great majority of studies investigating groove perception also point out that groove is associated with an experience of pleasure,'' based on studies that investigated groove perception. Thus, groove is considered to play an important role in the music-listening experience.
This study uses groove to evaluate the quality of the music-listening experience.
\section{Proposal}
This paper proposes a navigation method that stimulates the user's neck with musical vibrations modulated by the direction and distance to a destination.
\subsection{Haptic device}
\label{sec_pro_Haptic}
As mentioned in Section~\ref{rw_hapticNavi}, the conventional haptic navigation method uses a haptic device containing many small vibrators to utilize the spatial resolution of the skin.
These devices can certainly present locational information and musical vibrations but should be unsuitable for enhancing the groove of music because it is difficult for small vibrators to output low-frequency sound \cite{yamazaki2016tension}.
As for actuators other than vibrators, the possibility of their application to stimulate musical vibration is unknown.
However, in general, the temperature and driving force to deform the skin may not be suitable for driving musical vibration, considering the quick fluctuation of the music signal (several tens of milliseconds for the low-frequency part) and the amount of energy required to operate over a wide area.
\par
By contrast, our proposed necklace-type haptic device, Hapbeat (Fig.~\ref{fig_descPolar}(b)), can transmit vibrations of amplitude exceeding 10 $\mathrm{m/s^2}$ in a range of approximately 10--400 Hz over a wide area across the chest and neck and includes two built-in motors that drive the ribbon and stimulate the skin---a truly unique mechanism.
These characteristics create the impression that Hapbeat is suitable for stimulating musical vibrations and enhancing the perceived groove of music.
In addition, Hapbeat is designed for use during everyday travel. It is small (about 55$\times$58$\times$15 mm), lightweight (58.5 g), and easy to wear, making it suitable for the intended use in this paper: music listening while walking.
\par
In contrast, its spatial resolution is poor; Hapbeat users can distinguish vibrations only in two regions on both sides of the neck.
However, the upper limit of vibration output is large (i.e., the dynamic range of the vibration output is wide), and thus direction and distance information can be presented by modulating the amplitude of the input audio signals to the left and right motors \cite{yamazaki2019neck}.
Therefore, this paper uses Hapbeat as a device for stimulating musical vibrations.
Please refer to our previous paper, \cite{yamazaki2019neck}, for further details on Hapbeat.
\subsection{Modulation algorithm}
\label{sec_pro_algo}
\begin{figure}[hbt]
\begin{center}
\includegraphics[width=\linewidth]{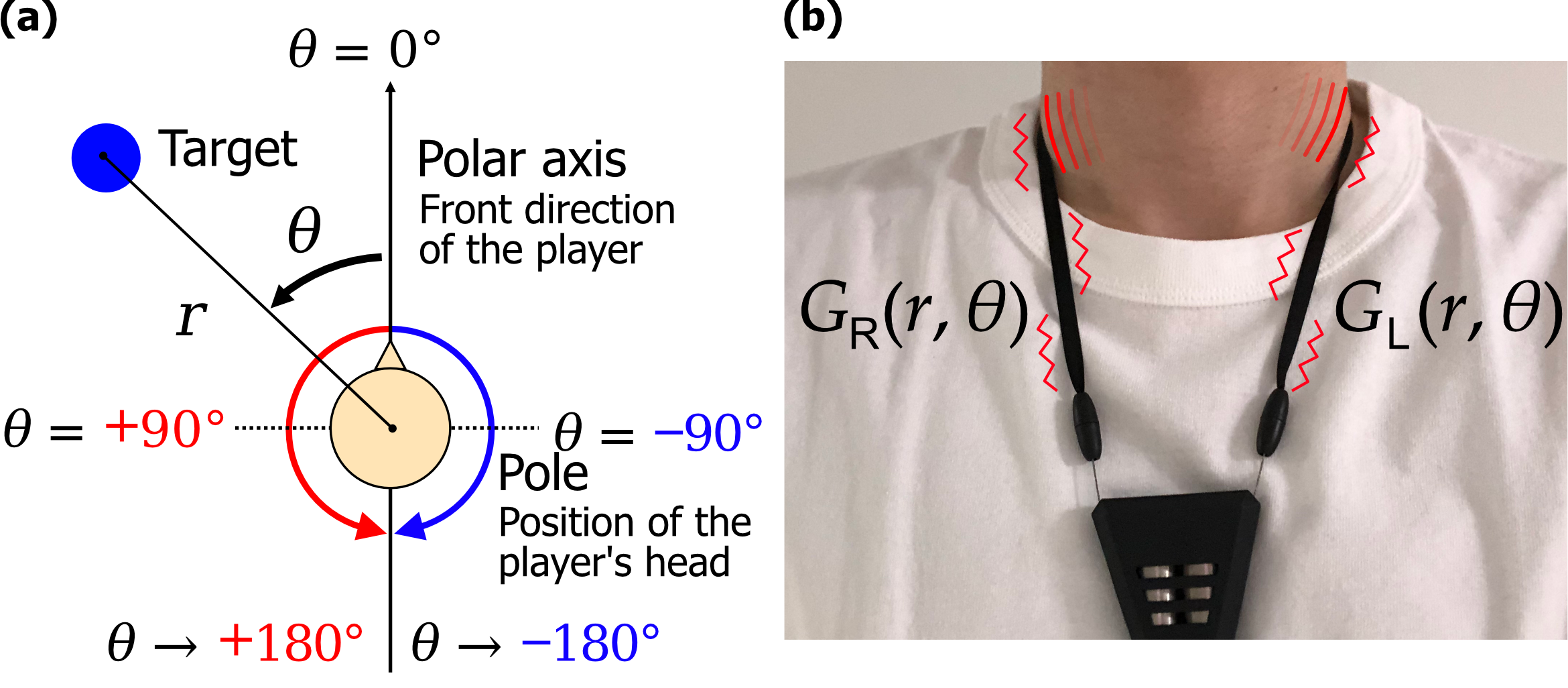}
\end{center}
\caption{
(a) Description of polar coordinates in our proposal.
(b) Appearance of Hapbeat.
In response to an input musical signal, built-in DC motors operate a ribbon on the neck, which transmits vibrations over a wide area around the sides of the neck.
The musical vibration based on the magnitude of $G_{\text{L}}( r,\theta )$ in Eq. \ref{eq_gain} is transmitted to the left side of the neck via the left side of the ribbon, while the same is done for the right side.
}
\label{fig_descPolar}
\end{figure}

Our method uses a polar coordinate system shown in Fig.~\ref{fig_descPolar}(a).
As in actual sound localization, a head tracking system constantly acquires the head orientation, and our method modulates musical vibration continuously according to the angle ($\theta$ in Fig.~\ref{fig_descPolar}(a)) toward the target (i.e., destination).
The directional information is presented by stereo balancing the entire musical vibration (Eqs. \ref{eq_dirLeft}, \ref{eq_dirRight}), with emphasis on clarity and with reference to the method of Jones \textit{et al.} \cite{jones2008ontrack}.
The distance information is presented by linearly increasing the total amplitude of musical vibration as the user approaches the target.
However, the equation is divided by cases according to the distance (Eq. \ref{eq_dist}) to prevent both sides of the vibration amplitude becoming zero when the distance is too large.
From the above, the final left and right vibration amplitudes are obtained by Eq. \ref{eq_gain}, where $G_{\text{L,R}}$ takes values from 0 to $C_{\text{Max}}$ depending on the distance and direction.
Fig.~\ref{fig_descExp}(a) shows a specific example of the modulation.
The musical signals, with an amplitude adjusted by the derived $G_{\text{L,R}}$, are input to the left and right motors of the Hapbeat---presented in Fig.~\ref{fig_descPolar}(b)---and converted into the ribbon movement, thereby stimulating both sides of the neck.
\begin{equation}
G_{\text{L,R}}( r,\theta ) =C_{\text{Max}} A_{\text{L,R}}( \theta ) A( r)
\label{eq_gain}
\end{equation}

\begin{equation}
A_{\text{L}}( \theta ) =\begin{cases}
0 & ( -180\degree \leq \theta \leq -90\degree ) \ \\
\frac{90\degree +\theta}{180\degree} & ( -90\degree \leq \theta \leq 90\degree )\\
1 & ( 90\degree \leq \theta \leq 180\degree )
\end{cases}
\label{eq_dirLeft}
\end{equation}

\begin{equation}
A_{\text{R}}( \theta ) =\begin{cases}
1 & ( -180\degree \leq \theta \leq -90\degree ) \ \\
\frac{90\degree -\theta}{180\degree} & ( -90\degree \leq \theta \leq 90\degree )\\
0 & ( 90\degree \leq \theta \leq 180\degree )
\end{cases}
\label{eq_dirRight}
\end{equation}

\begin{equation}
A( r) =\begin{cases}
1-\alpha r & \left( \ 0\leq r\leq \tfrac{1}{\alpha}\left( 1-\tfrac{C_{\text{Min}}}{C_{\text{Max}}}\right)\right) \ \\
\tfrac{C_{\text{Min}}}{C_{\text{Max}}} & \left(\tfrac{1}{\alpha}\left( 1-\tfrac{C_{\text{Min}}}{C_{\text{Max}}}\right) \leq r\right)
\end{cases}
\label{eq_dist}
\end{equation}
where $r$ is the distance and $\theta$ is the azimuthal angle (deg) between the user and target (as shown in Fig.~\ref{fig_descPolar}(a)), and where $\alpha$ is an arbitrary real number with $\alpha$, $C_{\text{Min}}$, and $C_{\text{Max}}$ determining the range of vibration modulation: the smaller the value of $\alpha$, the wider the vibration modulation range.
Note that this method only modulates musical vibrations (i.e., it does not affect the sound of the music).
\section{Evaluation}
The following experiments were conducted to test the hypotheses proposed in Section~\ref{sec_intro}.
Experiment 1 (Section~\ref{exp1_main}) consisted of conducting a questionnaire survey to examine the effect of stimulating musical vibration using Hapbeat on the music-listening experience while stepping to validate [H2].
Experiment 2 (Section~\ref{exp2_main}) evaluated the accuracy (deg) of users' ability to adjust their direction toward a target by stimulating modulated musical vibration to the neck in a virtual environment (VE) to validate [H1].
Experiment 3 (Section~\ref{exp3_main}) examined the ability of four different navigation methods by performing navigation tasks in VE and evaluating behavior logs and subjective impressions of ease of navigation and music-listening experience to validate [H1]--[H6].
\subsection{General experimental conditions}
\subsubsection{Participants}
Twenty-four participants ([male: female] = [18:6], age of [20s:30s:40s] = [21:1:2]) took part in the series of experiments conducted in this paper.
All participants were healthy and reported no abnormalities in sensory modalities.
Before starting the experiments, each participant signed a consent form based on human research ethics and received a payment of 1,500 JPY.
The duration of the experiment per person was about 1.5--2 hours.
Hereafter, each participant is numbered in experimental order and referred to as par 1, 2, ..., 24.
\subsubsection{Musical stimuli}
\label{sec4_music}
\begin{figure}[h]
\begin{center}
\includegraphics[width = 8cm]{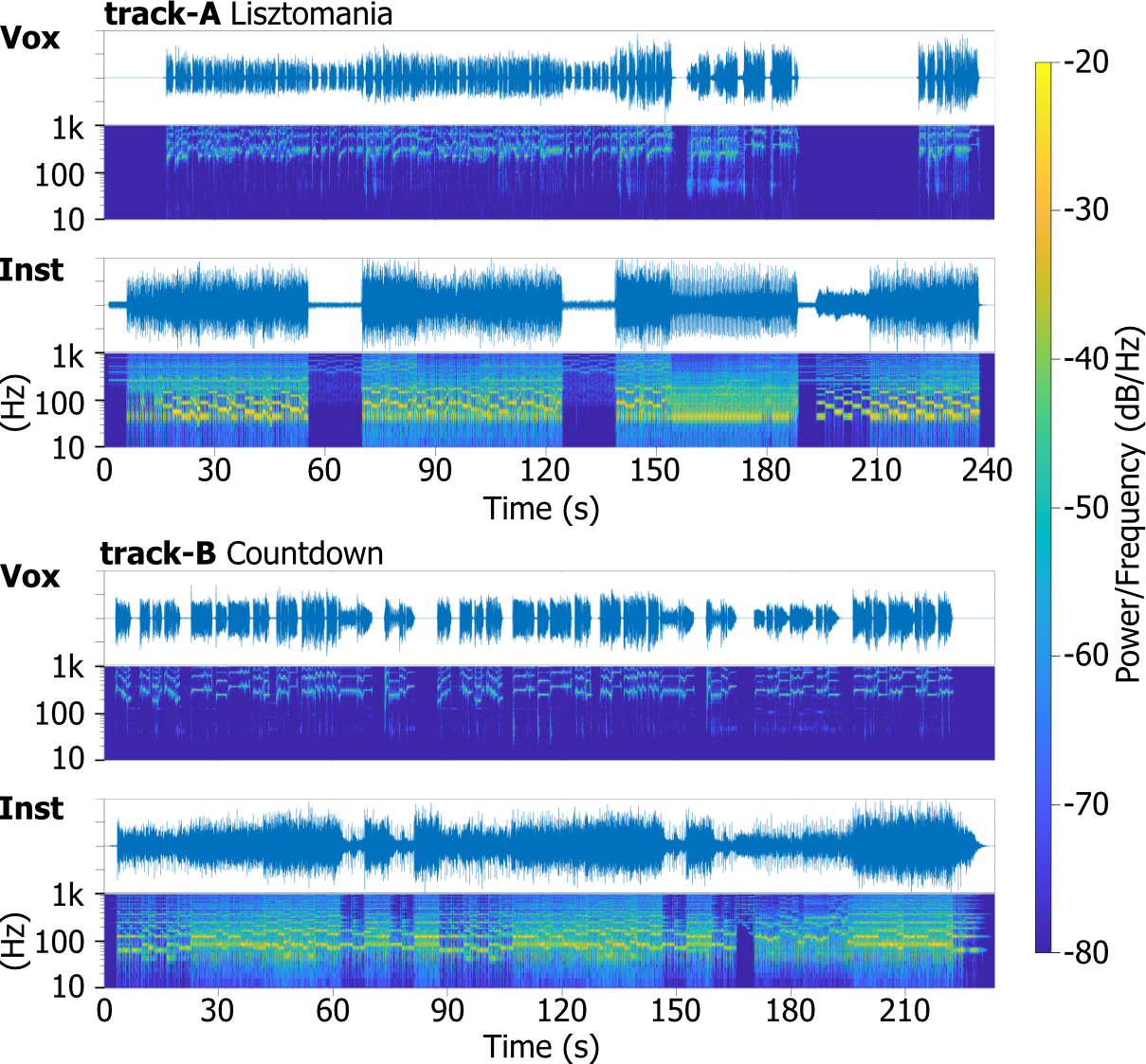}
\end{center}
\caption{
Waveforms and spectrograms of the music tracks (Vox and Inst) used as musical stimuli.
The vertical axis of the waveform indicates amplitude ($-1.0$--$1.0$), and the vertical axis of the spectrogram indicates frequency (Hz).
The graph's horizontal axis indicates the playback time of the piece(s).
The short-time Fourier transforms were calculated with 8192 samples using 50\% overlapping hamming windows.
}
\label{fig_descTracks}
\end{figure}

As musical stimuli, we used Phoenix's `Lisztomania' (track A) and `Countdown' (track B) available in multi-track format.
Of the multi-tracks, the tracks containing voices were grouped as Vox tracks, and the tracks containing other instruments and effects were grouped as Inst tracks, and their volumes were adjusted.
The waveforms and spectrograms are shown in Fig.~\ref{fig_descTracks}.
Both tracks have an orthodox instrumental structure that includes vocals, guitar, bass, and drums and contain enough bass to feel musical vibration with Hapbeat.
As loudness units, we used the units relative to full scale (LUFS, which is equivalent to LKFS in \cite{series2011algorithms}) defined in Recommendation ITU-R BS.1770 \cite{series2011algorithms}, which is generally used in loudness normalization of music tracks.
Comparing the two tracks, the loudness of the drum track and the bass track are similar in track A (drum: -18.9 LUFS, bass: -18.4 LUFS), while the loudness of the drum track is lower than that of bass track in track B (drum: -22.6 LUFS, bass: -19.7 LUFS).
Therefore, when played back using Hapbeat, track A emphasizes the rhythmic and percussive vibration of the bass drum to a greater extent than track B does, which emphasizes the continuous vibration of the bass.
Note that the loudness of each track was measured on the unedited tracks.
Note also that track A has a relatively long period of silence in both the Vox and Inst tracks compared to track B.
\par
The Vox and Inst tracks were normalized to the average LUFS of -14 and then composed as a Mix track, exported in mp3 format (192 kbps).
The loudness of each track was measured and normalized using a VST loudness meter plug-in (TBProAudio, dpMeter5) with Audacity (www.audacityteam.org) on the unedited tracks.
\subsubsection{Listening condition}
\label{exp_listeningCondition}
This paper uses Hapbeat to stimulate musical vibrations for the reasons described in Section~\ref{sec_pro_Haptic}.
Assuming use for when going out, compact and lightweight headphones (Audio-Technica Corp., ATH-S100) were used to play music, and no special measures were taken to block noise from the surroundings, such as using earmuffs.
In other words, the subjects possibly heard the audio noise from Hapbeat, but only one participant (par 5) answered that the audio noise was bothering them, asking about the noise after Experiment 1.
\par
To determine the audio volume during the experiment, participants first heard a 440 Hz sine wave sound at a loudness of 80 dBA ($\pm$2 dBA) played from a speaker.
Participants then adjusted the volume value (0--1) of the application used in the experiments so that the 440 Hz sine wave sound from the headphones was about the same loudness as that from the speaker.
Adjustments were executed in a quiet room (40$\pm$2 dBA) with the participant seated, and the speaker was placed about 1 m from the participant's head at about the same height (margin of error less than $\pm$20 cm).
Loudness measurements of the sound from the speaker were made with a sound level meter (Shenzhen Wintact Electronics Co., Ltd, GM1356) placed at the participant's head position before the adjustment was made by the participant.
The determined audio volume is common to all conditions in Experiments 1--3 for each participant.
\par
The vibration amplitude from Hapbeat was adjusted using an audio amplifier (Audio-Technica Corp., AT-HA2) so that the power consumption of one Hapbeat's motor was 1 W when an 80 Hz sine wave sound was applied at a volume value of 0.5 in the experimental application.
To investigate the degree of the transmitted vibration's amplitude, the acceleration of the ribbon was measured with each participant wearing Hapbeat using an accelerometer (NXP Semiconductors, MMA7361LC) and an oscilloscope (Tektronix Inc, MDO4024C) at a sampling rate of 10 kHz.
The accelerometer was attached to the ribbon with double-sided tape, and measurement points were determined visually in areas of greater curvature on both sides of the neck and in the middle of the nape.
The mean $\pm$SD ($\mathrm{m/s^2}$) amplitude of the measurements on the participants (n = 24) was [Left: Nape: Right] = [42.5$\pm$7.4:11.6$\pm$3.4:43.9$\pm$11.0].
All participants wore collarless clothing, and Hapbeat's ribbon was directly in contact with the skin on their neck.
\par 
Experiments 1 and 3 were conducted with the condition that the participants perform stepping, considering the effect of vibration during walking.
Participants wore sandals (Crocs Classic Clog) to unify foot conditions during the experiments.
The floor was concrete with 6 mm-thick tile carpets.
\subsubsection{Hardware}
The application used in the experiments was developed using the game engine software Unity (version 2020.3.22f1) and executed on a gaming laptop (CPU: AMD Ryzen7 4800H, RAM: 16 GB, GPU: Radeon RX 5500M).
As a video display, a head mounted display (HMD), Meta Quest2 (Meta Platforms, Inc.), was connected wirelessly to the laptop using Oculus Air Link, and an audio interface with four output channels (Behringer, UMC404HD) was used for audio and vibration signal output.
\subsubsection{Statistical hypothesis testing}
\label{sec_hypo}
For the corresponding data (differences in ratings between experimental conditions for the same participant), the Wilcoxon signed-rank test (pairwise comparison) was conducted with the null hypothesis of ``no difference in representative values between the two groups'' at the significance level $\alpha$ = 0.05.
For comparisons of data with no correspondence (data from different participants to be compared), the Wilcoxon rank-sum test (comparing two independent samples) was conducted with the null hypothesis of ``representative values between the two groups are the same'' at the significance level $\alpha$ = 0.05.
MATLAB R2022a was used as the statistical processing software.
\subsection{Experiment 1: Impression of music-listening experience with Hapbeat while stepping}
\label{exp1_main}
A questionnaire evaluation was conducted to investigate differences in the music-listening experience when listening to music with only headphones (headphone condition) and the combination of headphones and Hapbeat (Hapbeat condition).
Participants were arranged into two groups, and each group was assigned one track to listen to, either track A or track B.
The track not listened to here was used in Experiments 2 and 3.
Considering the effect of order, half of the participants listened in the headphone condition first and then in the Hapbeat condition, and the other half listened in the reverse order.
\subsubsection{Questionnaire}
\label{exp1_Questionnaire}
Oliver \textit{et al.} \cite{senn2020experience} investigated questions suitable for evaluating the groove feeling of music, and on that basis we developed a questionnaire comprising six questions.
The items were aligned in the order described as follows, and participants answered for each condition on a 7-point Likert scale.
The Likert scale had the following explanations for each value: 0---strongly disagree, 1---disagree, 2---slightly disagree, 3---neither/nor, 4---slightly agree, 5---agree, 6---strongly agree.
\begin{itemize}
\item
Q1: This music evokes the sensation of wanting to move some part of my body.
\item
Q2: This music is good for dancing.
\item
Q3: I cannot sit still while listening to this music.
\item
Q4: Listening to this music gives me pleasure.
\item
Q5: I like listening to this music.
\item
Q6: This music makes me feel good.
\end{itemize}
When answering for the Hapbeat condition, participants were instructed to consider `music' in the question as `listening experience,' including the sensation of musical vibration.
Participants were also instructed that they could edit the answer given in the first condition even after listening in the second condition.
\subsubsection{Procedure}
\label{exp1_procedure}
Participants first put on the headphones and Hapbeat and were asked to perform stepping in place while listening to the music.
Next, the participants listened to one assigned track in either the headphone or Hapbeat condition and then answered the questionnaire.
Participants then listened to the same music with the other condition and answered the questionnaire again.
\subsubsection{Result}
\label{exp1_result}
\begin{figure}[h]
\begin{center}
\includegraphics[width = \linewidth]{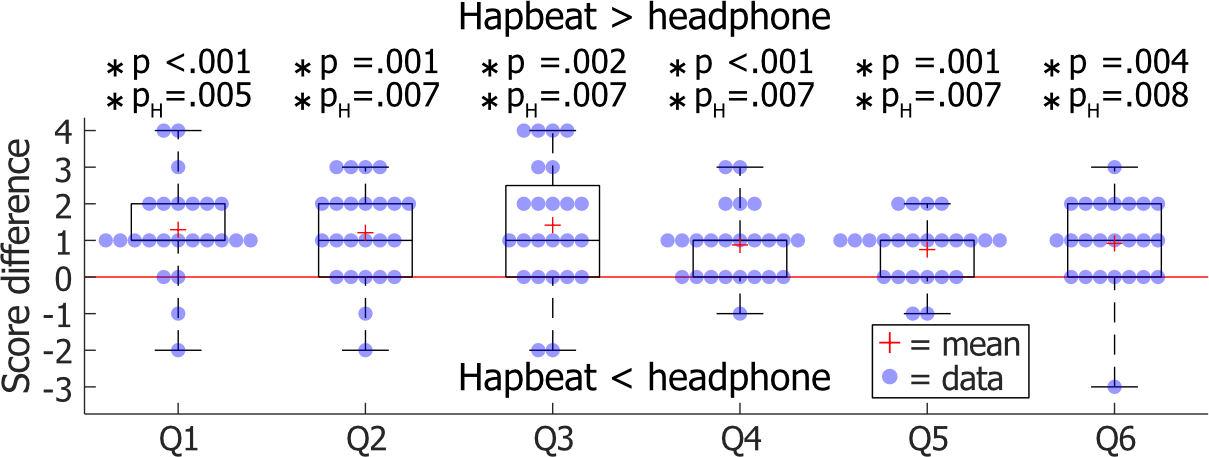}
\end{center}
\caption{
Score differences of questionnaire results in Experiment 1.
Score differences were calculated for each participant's score by subtracting the headphone condition score from the Hapbeat condition score.
For the derivation of p-values, see Section~\ref{sec_hypo}. Asterisks indicate significant differences.
The $\text{p}_\text{H}$-values are the p-values corrected by the Holm method (Section~\ref{discuss_h2}).
}
\label{fig_exp1result}
\end{figure}

The difference between the questionnaire results for the Hapbeat and headphone conditions for each participant is shown in Fig.~\ref{fig_exp1result}.
The Wilcoxon signed-rank test on the difference data (n = 24) for each question resulted in significant differences for all questions.
The Wilcoxon rank-sum tests for each question between groups that listened to track A and track B resulted in no significant differences for all questions.
\subsection{Experiment 2: Directional accuracy with modulated musical vibration stimuli to the neck}
\label{exp2_main}
\subsubsection{Evaluation target}
This experiment investigated the accuracy (deg) of users' ability to adjust their direction toward a target (hereafter referred to as front-detect-accuracy) by stimulating musical vibration modulated according to the algorithm (with $f(r)=1$) proposed in Section~\ref{sec_pro_algo}.
We focused on the front-detect-accuracy because we considered that localization of the direction to the target by the participant's head movement is important for navigation, based on previous studies described below.
Jones \textit{et al.} \cite{jones2008ontrack} reported that in the case of stereo panning modulation, the user tended to search for the correct direction (i.e., the direction where the left and right volume is equal) while changing the direction of a smartphone (corresponding to the position of the player's head in Fig.~\ref{fig_descPolar}(a))
Heller \textit{et al.} \cite{heller2009multi} also reported that turning the head was the key to navigation by ear (i.e., localizing audio sources).
One possible way of evaluating the direction presentation method would be to have participants describe the target’s location without changing their orientation.
However, based on the above reports, we judged that this localization method was not practical for navigation and thus did not evaluate it in this paper.
\subsubsection{Virtual environment for the experiment}
\label{exp2_ve}
\begin{figure}[h]
\begin{center}
\includegraphics[width = 7cm]{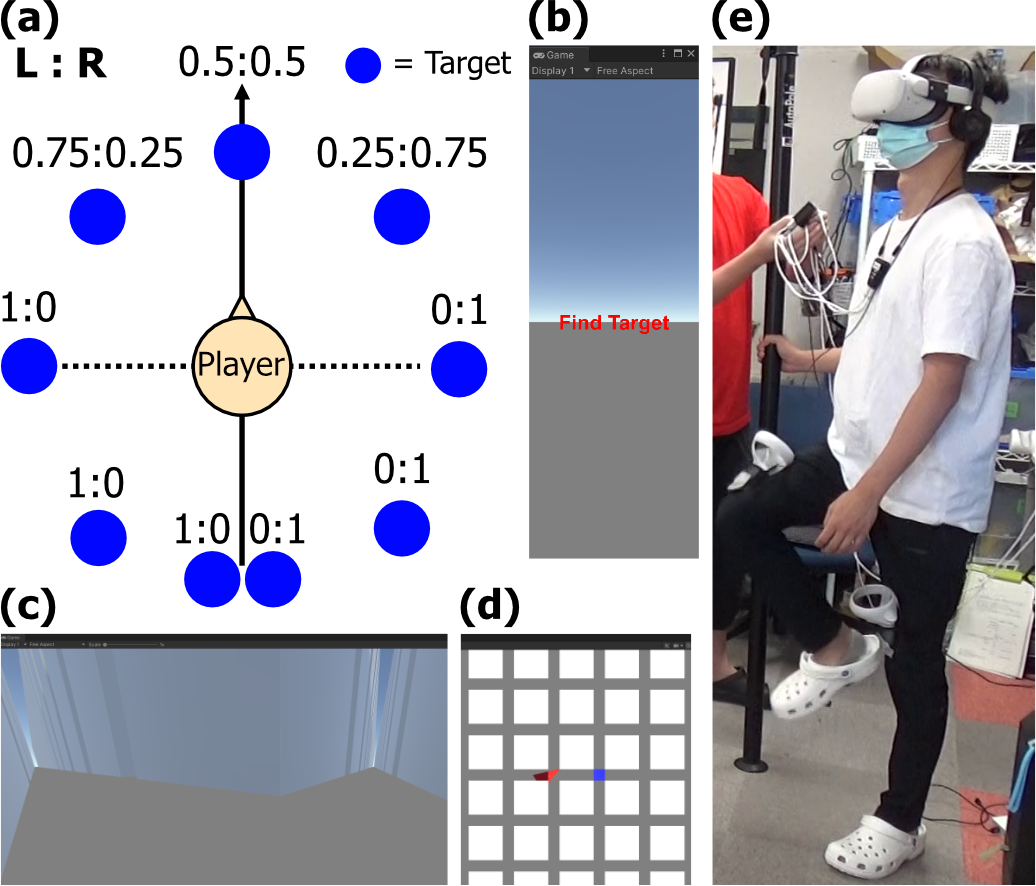}
\end{center}
\caption{
(a) Figure used in Experiment 2 to describe the proposed modulation algorithm for the participants.
The ratios indicate the ratio of vibration amplitude between the left and right side of the Hapbeat ribbon.
Please note that for clarity of explanation, the target's position in this figure does not show the target's placement in the experiment.
During the experiment, only one target appeared at any given time, randomly, continuously, and uniformly placed on the circumference of a circle at a radius of 2 m from the participant's position at the center.
(b) The VE in Experiment 2.
(c) The VE in Experiment 3.
(d) Overhead view during condition (c).
(e) Photo of a participant in Experiment 3.
}
\label{fig_descExp}
\end{figure}

This experiment was conducted in a VE using an HMD that can track the participant's head orientation to perform musical vibration modulation based on the participants' physical movements.
The VE is shown in Fig.~\ref{fig_descExp}(b).
The participant's front direction in the VE is synchronized with the orientation of the HMD; that is, with the direction the participant is facing in the real environment.
The experiment proceeded in the following three steps.
\begin{itemize}
\item
Step 1: The target was randomly, continuously, and uniformly placed on the circumference of a circle at a radius of 2 m from the participant's position at the center.
Note that Fig.~\ref{fig_descExp}(a) does not indicate the placement of the target in the experiment; rather, it shows the target's position in 45\textdegree~increments for illustrative purposes for the benefit of participants.
\item
Step 2: the participant turned their face (either by moving only their head or their whole body) to find the target by relying on modulated musical vibrations.
\item
Step 3: when the participant judged that the target was in front of them, they held down the controller’s grip button for one second to record their face direction.
\end{itemize}
The sequence of the above three steps is considered one `trial' in this experiment, and the trial was repeated until the track ended.
In other words, the total experiment time is unified across participants, but the total number of trials for each participant differs according to the response time for each trial.
Participants practiced the trial in tutorials before starting the experiment.
During the tutorials, the target was displayed as a blue sphere at the participant's eye level, and participants learned how the musical vibrations were modulated by checking the relationship between their face direction and the target.
\par
For each participant, the track listened to in Experiment 1 was used for the tutorial and the other track was used in the experiment as a music stimulus; that is, a participant who listened to track A in Experiment 1 was given the tutorial with track A and conducted the experiment with track B.
The answering time and error between the participant's direction and the direction to the target ($\theta$ in Fig.~\ref{fig_descPolar}(a)) were recorded when the participant answered and continuously at approximately 33 ms intervals during the experiment.
\subsubsection{Procedure}
The researcher first explained to the participants the modulation algorithm using Fig.~\ref{fig_descExp}(a) and the purpose of the experiment: to find the target in front of the participant, that is, turning their head to adjust the amplitude of vibration transmitted to both sides of the neck so it is the same.
Next, standing participants put on the HMD and were given the tutorial.
After ensuring that the participant fully understood how the musical vibration was modulated and how the trials described above would be conducted, the researcher instructed them to focus on accuracy rather than response speed and started the experiment.
The experiment lasted until the end of the track, with participants completing the trials repeatedly at their own pace.
\subsubsection{Result}
\label{exp2_result}
\begin{figure}[h]
\begin{center}
\includegraphics[width = \linewidth]{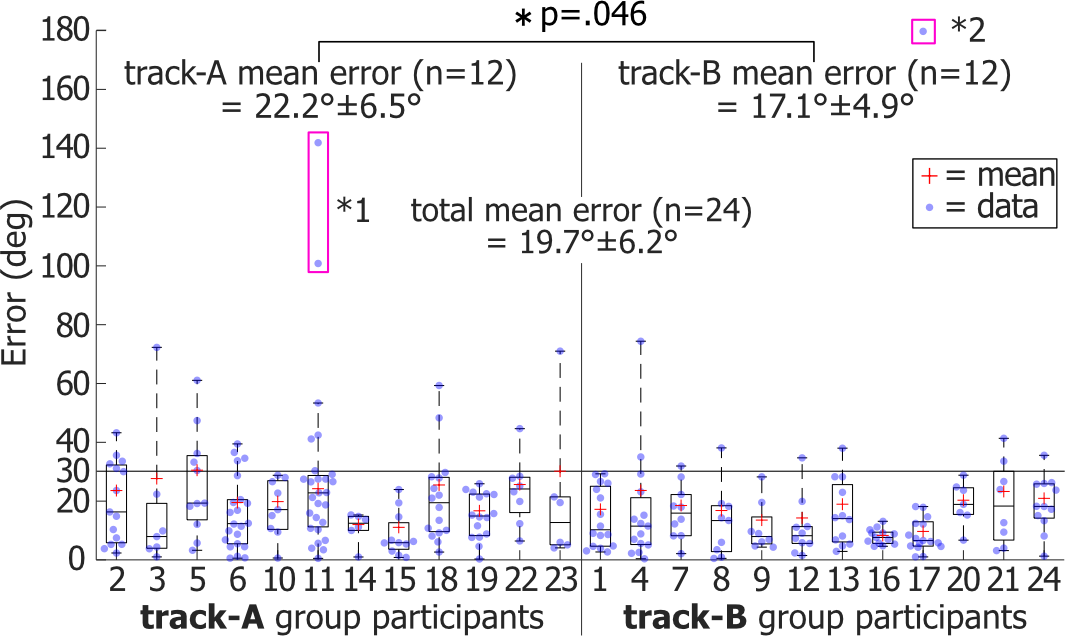}
\end{center}
\caption{
Result of Experiment 2.
Data points framed in magenta are excluded from calculation of the mean.
For the derivation of p-values, see Section~\ref{sec_hypo}. Asterisk beside p indicate significant differences.
Note that the data framed in magenta in the figure are excluded from the mean calculation for the following reasons:
*1: We judged the trials to be an operational error because the participant was upright and immobile, and their responses were recorded within a short interval (less than 3 s).
*2: We judged that the participant mistakenly thought the target directly behind them was in front because they moved their head from side to side just before recording.
}
\label{fig_result_exp2}
\end{figure}

All recorded data and the mean of each participant's mean error (deg) are shown in Fig \ref{fig_result_exp2}.
For each participant, the percentage of outcomes with an error margin below 30\textdegree~was calculated (only *1, a manipulation error, was excluded) in order to compare the results with those presented in the work of Heller \textit{et al.} \cite{heller2018navigatone}.
The mean of the percentages for all participants was 89\%; for the track A group it was 85\%  and for the track B group 92\%.
The mean value of mean time spent to complete each trial for all participants was 21.1$\pm$8.3 s; for the track A group it was 22.0$\pm$10.2 s and for the track B group 20.3$\pm$6.3 s.
There was no significant difference between the tracks (p = .167 for the percentage and p = .795 for the time).
\subsection{Experiment 3: Evaluating navigation ability while walking}
\label{exp3_main}
\subsubsection{Navigation conditions}
The experiment was conducted under the four types of location information navigation conditions described in table \ref{tab_conditions} and below.
\begin{itemize}
\item
NavigaTone (NT): the condition that played music whose Vox track was localized by the direction to the target.
The method is based on previous studies \cite{heller2018navigatone, heller2020attracktion} using the Resonance Audio framework \cite{gorzel2019efficient} to perform spatial audio rendering.
\item
NT\&Hap: the condition that stimulated the unmodulated musical vibration to the participant in addition to the NT.
In other words, location information is presented only by the localized Vox track, as in the NT.
\item
HapDir: the condition that stimulated the musical vibration, modulated according to the algorithm proposed in Section~\ref{sec_pro_algo}, to the participant and simultaneously played music without modulation.
However, no modulation by distance was performed (Eq. \ref{eq_dist} $A(r) = 1$).
\item
HapDirDist: the condition that presented distance information by modulating the total amplitude of musical vibration as shown in Eq. \ref{eq_dist}.
We set $C_{\text{Max}}$ = 1 and $C_{\text{Min}}$ = 0.2 for all participants, and other conditions are same as for HapDir.
\end{itemize}
\begin{table}[h]
\centering
\caption{
Summary of modulation under different navigation conditions
}
\label{tab_conditions}
\begin{tabular}{lll}
Conditions & Audio       & Haptic         \\ \hline
NT         & Mod Vox Dir & None             \\
NT\&Hap    & Mod Vox Dir & Unmod All        \\
HapDir     & Unmod All   & Mod All Dir      \\
HapDirDist & Unmod All   & Mod All Dir+Dist
\end{tabular}
\end{table}
Each condition is intended to verify the hypotheses [H1]--[H6].
To verify [H1] and [H4], the NT and HapDir were compared; to verify [H2] and [H3], the NT and NT\&Hap were compared; to verify [H5] and [H6], the HapDir and HapDirDist were compared.
Of these, we prioritize the validation of [H1], the main purpose of this paper, and considered the effects of order in the NT and HapDir conditions, dividing participants into two groups: those who undertook the trial in the order NT, HapDir, NT\&Hap, and HapDirDist, and those who undertook the trial in the order HapDir, NT, HapDirDist, and NT\&Hap.
\subsubsection{Virtual environment for the experiment}
\begin{figure}[hbt]
\begin{center}
\includegraphics[width = \linewidth]{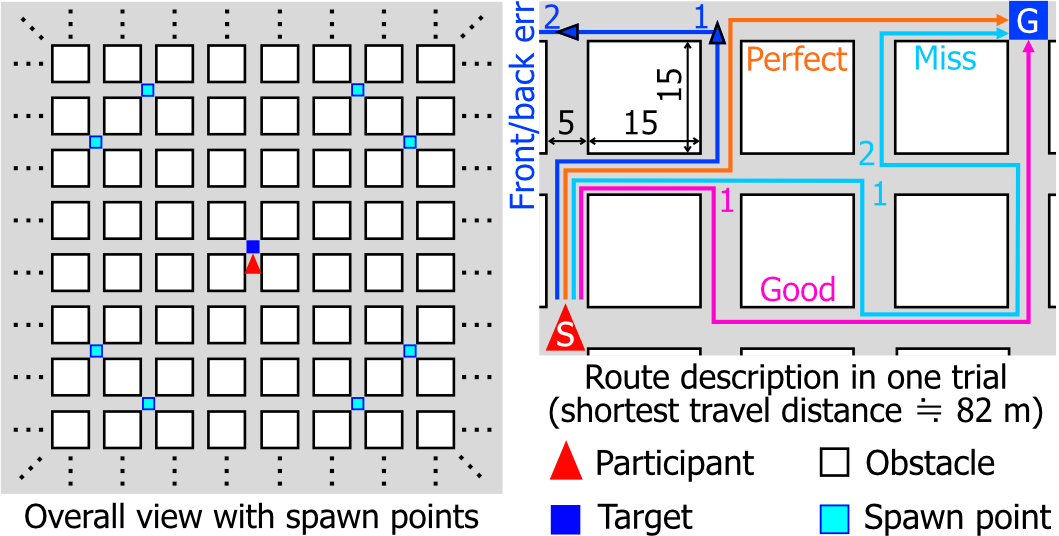}
\end{center}
\caption{
VE description in Experiment 3.
Left: When a participant arrives at a target, the target respawns at a random location from the next eight spawn points.
Right: An example of a route type in one trial.
The colored lines and texts indicate [Perfect: Good: Miss: Front/Back err] = [orange: magenta: cyan: blue], and colored numbers indicate the number of times the participant made a mistake at the crossroads.
The shortest travel distance was calculated considering diagonal travel.
}
\label{fig_naviRoute}
\end{figure}

The experiment was performed in the VE using an HMD as in Section~\ref{exp2_main}.
The VE view and photo of the participant are shown in Figs. \ref{fig_descExp}(c)--(e), and the detail of the VE is shown in Fig.~\ref{fig_naviRoute}.
The experiment defines a `trial' as a participant attempting to reach the target in the VE by walking.
Once the experiment begins, the target is spawned at a random position from among the light-blue square locations shown in Fig.~\ref{fig_naviRoute}, based on the origin of the VE (participant's initial position).
Contact between the participant and the target is regarded as the target having been reached, and then the system plays a cue: a ping sound effect in NT and NT\&Hap, and a vibration of one-second of a 10 Hz sine wave in HapDir and HapDirDist.
The target is then randomly respawned as described above.
\par
Participants can move forward by stepping and can change the direction by facing in the direction they wish to go.
To detect stepping movements, the HMD's controllers were attached to the thigh above the knee using a rubber belt (Fig.~\ref{fig_descExp}(e)).
The algorithm for detecting walking motion was as follows: the movement of the leg was considered as one step in the VE when the controller's height exceeded a certain value ($+10$ cm in this paper), based on the controller's position when standing, and then the height fell below that value.
For each step, the participant's position in the VE was advanced 1.17 m in 0.7 s (approximately 100 m/min).
\par 
In the tutorial, participants first reached the visualized target (shown as a blue rectangular pillar) several times in an environment without obstacles in order to practice walking in the VE.
Next, the participants practiced walking in an environment with obstacles, as shown in Fig.~\ref{fig_naviRoute}.
Finally, the participants experienced and learned the location presentation method until they fully understood how they were guided by the modulated sound or vibration.
\par
The track used for the tutorial and the experiment was the same as in Experiment 2 (described in Section~\ref{exp2_ve}).
The participant's coordinates and elapsed time were recorded continuously at approximately 33 ms intervals during the experiment.
\subsubsection{Behavior evaluation}
The degree of achievement of navigation conditions was evaluated by the group ratio grouped by the following criteria.
We define `correct route' here as the shortest route to reach the target.
\begin{itemize}
\item
Perfect: A trial where the participant chooses the correct route at all crossroads (e.g., the orange line in Fig.~\ref{fig_naviRoute}, right).
\item
Good: A trial where the participant chooses the wrong route one time (e.g., the magenta line in Fig.~\ref{fig_naviRoute}, right).
\item
Miss: A trial where the participant chooses the wrong route twice or more (e.g., the cyan line in Fig.~\ref{fig_naviRoute}, right).
\end{itemize}
The completion time for Perfect trials was measured and evaluated.
To measure the extent to which participants became lost, we calculated the mean distance of the Miss trials.
We also tabulated the results, defining ``failure to distinguish between the front and back'' as two or more consecutive trials of making the wrong choice with the target behind the participant at a crossroads (e.g., the blue line in Fig.~\ref{fig_naviRoute}, right) and referred to as ``Front/Back err''.
\subsubsection{Subjective evaluation}
For the subjective evaluation, we conducted a questionnaire using the 7-point Likert scale for the following items.
The format is the same as in Section~\ref{exp1_Questionnaire}, with the adjectives used instead of `agree' or `disagree' shown in parentheses.
\begin{itemize}
\item
Q1: Was it easy to find the target? (difficult--easy)
\item
Q2: How did the music modulation have an impact (affect) on your musical experience compared to simply listening to music? (negative--positive)
\end{itemize}
Participants completed the questionnaire for each condition at the end of each experiment, and they were able to modify their evaluations after conducting trials under other conditions.
Thus, the questionnaire evaluation can be compared across the conditions.
\subsubsection{Procedure}
First, participants attached the controllers to their legs using rubber belts.
Next, participants put on the HMD and were given the tutorial before each condition.
After the researcher confirmed that the participant fully understood the navigation method under the condition, the participants started the experiment.
During the experiment, the participant held on to a fixed pole (Manfrotto, autopole) with one hand to prevent them from moving by stepping.
In each condition, the participant conducted the trial three times, and then they removed the HMD and answered the questionnaire.
After the experiment was completed in all conditions and the participant had answered the questionnaires, the researcher interpreted the results of the questionnaire and communicated their interpretation to the participants to check whether the results correctly reflected their subjective impressions.
The researcher also asked them questions about the reasons for their evaluation differences between the conditions.
During this interview, the researcher focused only on eliciting the participants' thoughts and tried not to induce them; participants were free to modify their scores.
\subsubsection{Result}
\label{exp3_result}
The ratio of navigation accomplishment trials for each presentation condition is shown in Table \ref{tab_exp3result}.
Participants could reach the target only through the navigation conditions for each trial.
The Perfect, Good, and Miss rows in the table are tabulations for all participants, while the others are tabulations of the percentage of Perfect trials for each group.
The number in parentheses next to the number indicates the trials counted.
The mean arrival times(s) for Perfect trials were [NT: NT\&Hap: HapDir: HapDirDist] = [89:82:100:90], with no significant differences between navigation conditions.
In addition, no significant differences were found comparing the mean arrival times(s) for Perfect trials between track A and track B groups for all navigation conditions.
\par
\begin{figure}[hbt]
\begin{center}
\includegraphics[width = \linewidth]{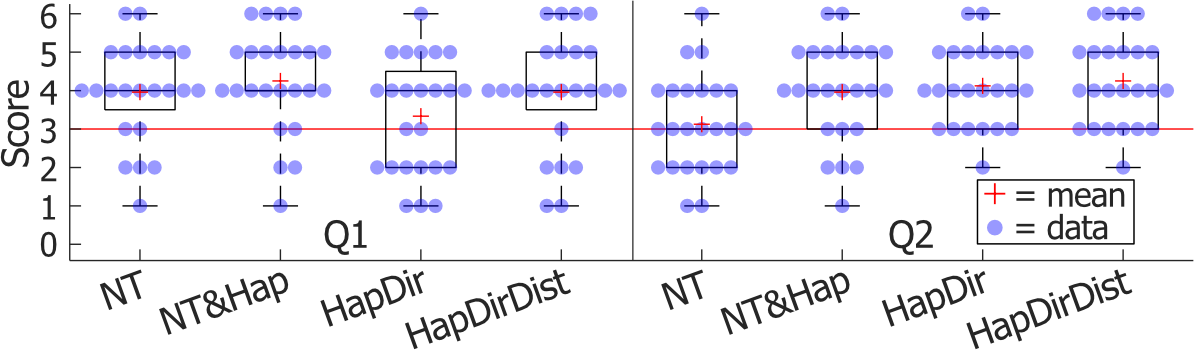}
\end{center}
\caption{
Result of the questionnaire in Experiment 3.
}
\label{fig_result_exp3_raw}
\end{figure}
\begin{figure}[h]
\begin{center}
\includegraphics[width = \linewidth]{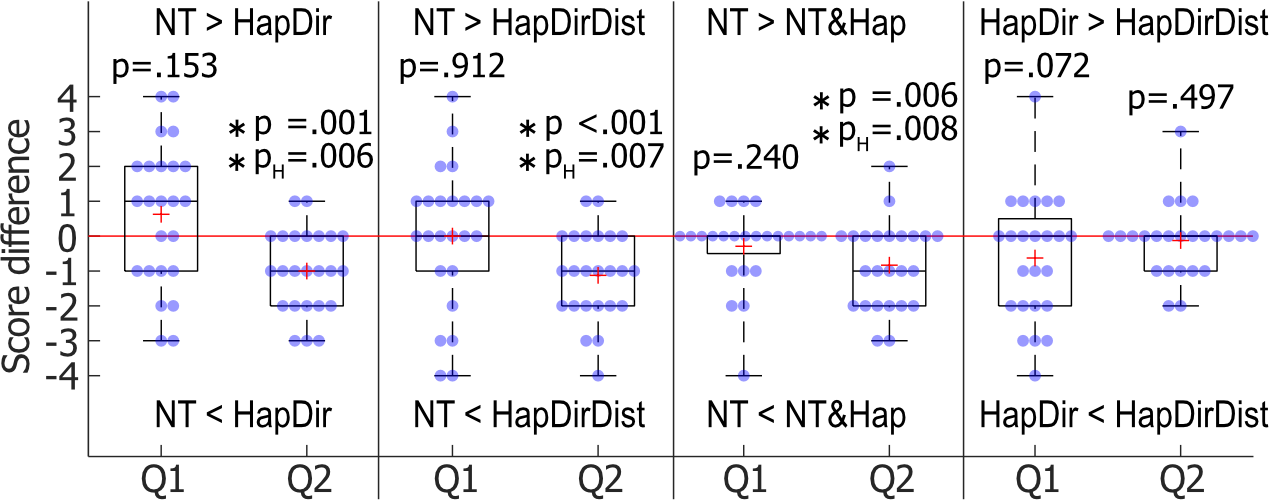}
\end{center}
\caption{
Score differences of questionnaire results in Experiment 3.
Score differences are calculated for each participant's score between the pair of navigation conditions in the figure.
For the derivation of p-values, see Section~\ref{sec_hypo}. Asterisks indicate significant differences.
The $\text{p}_\text{H}$-values are the p-values corrected by the Holm method (Section~\ref{discuss_h2}).
}
\label{fig_result_exp3_diff}
\end{figure}

The mean travel distances (m) for the Miss trials were [NT: NT\&Hap: HapDir: HapDirDist] = [162:165:196:283]
and the number of Front/Back errs was [NT: NT\&Hap: HapDir: HapDirDist] = [0:1:3:7].
\begin{table}[h]
\centering
\caption{
Group ratio of navigation task
}
\label{tab_exp3result}
\begin{tabular}{ccccc}
Group                  & NT   & NT\&Hap & HapDir & HapDirDist \\ \hline
Perfect                & 0.89 (63) & 0.93 (67)    & 0.79 (56)   & 0.81 (58)       \\
Good                   & 0.10 (7) & 0.06 (4)    & 0.11 (8)   & 0.07 (5)       \\
Miss                   & 0.01 (1) & 0.01 (1)    & 0.10 (7)   & 0.13 (9)       \\  \hline
track A                & 0.83 (30) & 0.89 (32)    & 0.74 (26)   & 0.75 (27)       \\
track B                & 0.94 (33) & 0.97 (35)    & 0.83 (30)   & 0.86 (31)       \\
\end{tabular}
\end{table}
The questionnaire results are shown in Fig.~\ref{fig_result_exp3_raw} and the score differences between conditions are shown in Fig.~\ref{fig_result_exp3_diff}.
\section{Discussion}
The results of the experiment indicate that stimulating modulated musical vibration is a practical way of guiding the participants and enhancing their music-listening experience.
We discuss the results that verify the hypotheses made in Section~\ref{sec_intro} below.
\subsection{H1: Modulating the stereo balance of musical vibrations can convey enough information for navigation.}
\label{discuss_h1}
We consider that [H1] is supported.
The results of Experiment 2 (Section~\ref{exp2_result}) show that the proposed method allows participants to identify the target direction with an accuracy of approximately 20\textdegree.
The percentage of errors less than 30\textdegree~was 89\%, which is comparable to the accuracy of NavigaTone proposed by Heller \textit{et al.} \cite{heller2018navigatone} (users can judge the direction to the target within 30\textdegree~error in 86\% of cases).
In addition, Williamson \textit{et al.} \cite{williamson2010social} reported that 60\textdegree~accuracy is sufficient for practical navigation.
In the navigation task of HapDir and HapDirDist in Experiment 3, participants were able to reach the target in all trials, and participants took the shortest route in about 80\% of cases.
Therefore, we assume that stimulating modulated musical vibration can present sufficient information for navigation, although it is undeniably more difficult to understand than NavigaTone judging from the results of NT and NT\&Hap.
\par
The proposed method does, however, have some issues.
About a third of participants rated HapDir as difficult to understand (score 0--2 for Q1 in Fig.~\ref{fig_result_exp3_raw}).
For this reason, par 7 commented, ``HapDir require[d] more concentration to localize the target than NT,'' and par 21 commented, ``NT's locating resolution [was] higher than that of HapDir.''
One of the reasons the participants found it difficult in Experiment 3, even though they were able to identify the direction to the target with sufficient accuracy in Experiment 2, could be that the obstacles in the experimental environment are aligned in a grid, as shown in Fig.~\ref{fig_naviRoute}.
When participants approached a crossroads, they tended to stop and quickly turn their heads in the direction of the pathways, that is, in 90\textdegree~increments, to search for the correct direction.
musical vibrations from bass sounds are not continuous like a sine wave, and the beginning of a sound, such as when a performer hits a drum or plucks a bass guitar string, will be emphasized.
This characteristic of musical vibration may cause the participants to perceive it discretely.
Therefore, we assume that participants find it difficult to identify the target's direction because they cannot continuously feel the change in musical vibration from both sides of their neck when they turn their face quickly and make larger directional movements.
One possible method to solve this problem is to mix a sine wave with the music signal of the vibration source: the frequency and amplitude of such a sine wave should be adjusted so as not to interfere with music listening.
\par
The participants took the shortest route in almost all trials in NT and NT\&Hap, even though they turned their faces quickly and made larger movements at crossroads.
One possible reason is that the vocal sounds had a long duration, and participants were able to perceive the sound changes continuously.
In addition, researchers have shown that the accuracy of detecting sound localization using head movements is less dependent on the speed of the head movements \cite{honda2016detection}, so we assume that participants were able to localize the target's position with the same accuracy as when using slower head movements.
\par
Four participants commented that the proposed algorithm made it difficult to distinguish between front and back; in fact, we observed several cases in our experiments in which participants seemed to have mistakenly distinguished between front and back.
One reason could be that Hapbeat's band is not separated, so even if only one side of the motor is activated, Hapbeat cannot completely eliminate the vibration transmitted to the other side.
Another reason could be that when the target is directly behind the participant, the left-right vibration switches in small head movements meant that it was perceived as if both sides of the ribbon were vibrating.
This is a serious problem: if the participant loses the distinction between front and back, they get lost and the arrival time significantly increases (819 seconds in the longest case).
Therefore, it is necessary to improve the algorithm so that front and back can be reliably distinguished; for example, stop vibration when the target is behind the user.
\subsection{H2: Stimulating musical vibration while walking enhances the music-listening experience.}
\label{discuss_h2}
We consider [H2] to be supported.
To verify [H2], we defined the null hypothesis family in a total of nine items: six from Q1--Q6 in Experiment 1 comparing Hapbeat and headphone conditions and three from Q2 in Experiment 3 comparing NT with the other three conditions.
The p-values were corrected using the Holm method \cite{holm1979simple} while taking into account the multiple testing problem, and significant differences were found for all items simultaneously, as shown in Fig.~\ref{fig_exp1result} and Fig.~\ref{fig_result_exp3_diff}.
From this we can see that, in all cases, the participants preferred the music-listening experience with musical vibration.
Thus, the results demonstrate that musical vibration improves the music-listening experience, even when the listener is distracted during walking and navigation, when the foot is stimulated when stepping, and when there is modulation of the amplitude of musical vibrations.
A common response in Experiments 1 and 2 was for participants to positively report that the musical vibrations ``increase[d] the power and intensity of music'' and ``ma[de] it easier for [them] to concentrate and immerse [themselves] in the music.''
\subsection{H3: Stimulating musical vibration while walking does not interfere with the previously proposed navigation method that modulates musical sound.}
\label{discuss_h3}
We consider [H3] to be supported.
Table \ref{tab_exp3result} shows that NT and NT\&Hap have almost the same rate of Perfect trials and arrival times, with the difference between NT and NT\&Hap in Q1 of Fig.~\ref {fig_result_exp3_diff} implying that the stimulated musical vibrations would not interfere with navigation, even after considering learning effects.
Therefore, combining musical vibrations with existing navigation methods based on music modulation is likely to be suitable in terms of presenting direction.
\par
However, musical vibrations may impact users' auditory attention to music.
Five participants who rated NT\&Hap higher than NT (par 4, 8, 16, 17, 24) noted that ``the vibration helped [them] to hear the voice clearly.''
In contrast, all participants who rated NT\&Hap lower than NT (par 6, 9, 11, 20) observed that ``the vibration distracted [them] from focusing on the voice.''
These impressions appear to reflect the result of previous study \cite{merchel2009hearing}, which demonstrated that stimulating vibrations synchronized with audio cues increase the perception of auditory loudness but differ in terms of the audibility of the Vox track.
The musical vibration from Hapbeat is supposed to emphasize the low-frequency range, that is, the drum and bass tracks, meaning that the vibration enables participants to more easily listen to the bass part of the music.
Therefore, we assume that some participants could perceive the music as described in par 24---``feeling the musical vibration separated the bass part of the music from the voice, and I felt that the voice sounded extra emphasized''---while others felt that the emphasized bass sound masked the sound of the Vox track.
Thus, the type of music track stimulated as a musical vibration may affect the user's auditory attention to music. To clarify this, further investigation will be required.
\subsection{H4: Navigating by modulating musical vibration is preferred as a music-listening experience over modulating music sound.}
\label{discuss_h4}
We consider [H4] to be supported.
Fig.~\ref{fig_result_exp3_diff} shows that participants scored HapDir significantly higher than NT in Q2.
Regarding the absolute ratings of Q2 shown in Fig.~\ref{fig_result_exp3_raw}, only one person gave a negative score for HapDir.
Five participants (par 6, 11, 14, 15, and 18) who shared their impressions of HapDir concerning the change in intensity of the left-right vibration all stated that it had no negative effect on their music-listening experience.
By contrast, eight participants (par 3, 4, 6, 11, 12, 14, 15, and 21) were negative about the listening experience in NT, with all of them clearly stating that the localization of the Vox track negatively affected their music-listening experience.
This indicates that a certain number of users will feel that modulating the voice track detracts from the listening experience.
\par
Nevertheless, this does not mean that NT typically impacts the music-listening experience in a negative way.
Indeed, nine participants (par 1, 2, 8, 9, 13, 16, 17, 18, and 25) gave NT a positive score, stating that ``the vocal localization gave a realistic feeling of a live performance,'' which suggests that vocal localization may have a positive effect for some users.
Meanwhile, four of the eight participants who scored NT negatively (par 3, 4, 14, and 21) gave a positive score (score 4--6 for Q2 in Fig.~\ref{fig_result_exp3_raw}) to NT\&Hap, stating that ``the musical vibration reduced the discomfort caused by vocal localization.''
This may be because the musical vibration focused participants' auditory awareness of the low frequencies, which in turn may have reduced their awareness of the vocal part, as mentioned in Section~\ref{discuss_h3}.
This indicates that, for some users, stimulating musical vibrations can reduce the unnaturalness of voice modulation.
Although further research is needed to clarify the relationship between user attributes and users’ perceptions of the modulation method, the proposed method certainly contributes to increasing the number of viable options for each user.
\subsection{H5: Presenting distance information by modulating the amplitude of the musical vibrations  makes it easier for users to understand the navigation.}
\label{discuss_h5}
We consider that further research is needed to verify [H5].
Comparing HapDir and HapDirDist, the result of group proportions shown in table \ref{tab_exp3result} was almost the same, and no significant differences were found in Q1 shown in Fig.~\ref{fig_result_exp3_diff}.
The average travel distance and the number of Front/Back err in Miss group were more in HapDirDist.
From the above, the results cannot support [H5].
However, for the following reasons, we claim that the proposed method has successfully presented distance information, and its effectiveness needs to be properly verified.
\par
On the negative side, seven participants (par 7, 8, 9, 10, 16, 19, 24) commented ``the vibration was weak when far from the target, making it difficult to determine the direction and front/back.''
Combining this comment with the fact that the average travel distance and the number of Front/Back err in Miss group were more in HapDirDist than in HapDir, we assume that those participants could not clearly identify the direction when the vibration was too weak at far from the target (e.g., at the beginning of a trial), making it more difficult to find the correct direction when they lost.
Although perceptual thresholds for tactile sensation have large individual differences \cite{aevarsson2022vibrotactile}, we have not adjusted vibration intensity for each participant in this study.
Therefore, ignoring the tactile sensitivity of each participant likely influenced the results of HapDirDist more than the modulation method.
\par
In contrast, the number of participants who answered that HapDirDist was difficult to understand (scored 0--2 for Q1 in Fig.~\ref{fig_result_exp3_raw}) was three compared to eight participants for HapDir.
Positive comments about HapDirDist obtained from participants included the following: ``The distance information help[ed] [them] to imagine the target's position'' (par 4, 5, 6, 12, 15); ``The distance information g[ave] [them] confidence that [they were] going in the right direction'' (par 4, 5, 9, 13, 17, 18, 19, 20, 21).
Although there were no significant differences, the results also showed that the participants could reach the destination faster in trials of Perfect group and felt easier to understand from the Q1 result in Fig.~\ref{fig_result_exp3_diff} under HapDirDist compared to HapDir.
This suggests that the proposed method succeeds in expressing the distance and may contribute to assuring participants of approaching the destination, who could determine the direction even with the minimum vibration intensity in the experiment.
The arrival time can be an effective indicator for evaluating the assurance; for example, the more confident participant possibly walked faster and spent less time selecting the way at the crossroads.
However, the arrival time was mostly spent with constant walking speeds on the pathways in the experiment, so the degree of assurance could not be properly evaluated.
Therefore, further research is needed to investigate the contribution of presenting distance information by (1) introducing a calibration procedure to determine the minimum vibration intensity according to each participant's tactile perception and (2) redesigning the experiment systems and evaluation items to measure the degree of assurance appropriately.
\subsection{H6: Presenting distance information by modulating the amplitude of the musical vibrations negatively impacts the music-listening experience.}
\label{discuss_h6}
We consider that [H6] is not supported.
From Figs. \ref{fig_result_exp3_raw} and \ref{fig_result_exp3_diff}, it can be seen that the participants scored HapDir and HapDirDist almost equally (no significant difference) on Q2, with only one participant giving each one a negative score.
This can be attributed to no participants commenting that changing the intensity of the musical vibration negatively affects the listening experience, as described in Section~\ref{discuss_h4}.
As a positive for HapDirDist, six participants (par 10, 12, 13, 17, 19, 20) commented that ``the increasing vibration ma[de] [them] feel as if [they were] getting closer to an imaginary performer which evokes a live music feeling, and that was fun.''
Thus, for some users, changing the intensity of the musical vibration can enhance the listening experience.
In contrast, two participants (par 11, 22) rated HapDirDist lower than HapDir due to lower intensity, confirming that [H6] can occur in a few users.
These results indicate that in many cases, modulating the amplitude of musical vibration to present distance information is unlikely to affect the music-listening experience adversely.
\subsection{Comparing music tracks}
\label{discuss_track}
The results of the experiments suggest that the characteristics of the music tracks may affect the clarity of navigation, although there were no significant differences in arrival time.
The participants in the track B group had significantly fewer errors in frontal identification in Experiment 2, and the percentage of Perfect trials in Experiment 3 was also higher.
As shown in Fig.~\ref{fig_descTracks}, track B has a shorter period of silence and a louder bass track compared to track A, resulting in the participants being able to feel the musical vibration more continuously.
This may have made it easier for participants to feel changes in vibration intensity modulated by the proposed method, making track B more suitable for navigation.
\par
From Table \ref{tab_exp3result}, track B participants performed better than track A participants in NT and NT\&Hap, both using navigation by voice.
This may be due to the shorter silence time in the Vox track for track B than in track A, and participants were able to perceive the target's location relatively further away.
Therefore, a music track that has a longer period of silence could be unsuitable for navigation.
For such music, combining NavigaTone and the proposed method will greatly improve navigation performance with relatively little disruption to the listening experience.
In any case, we assume that the proposed method applies to many music tracks, including bass instruments, because participants always reached the target regardless of the use of different tracks.
\subsection{Limitations}
The impact of the difference between the VE used in this paper and the real environment cannot be ignored.
First, the experimental system is very different from real-world walking.
Participants were reminded to move their knees up and down more significantly than they would in a real walking motion so the system could detect their actions.
Participants always move forward in the direction of their face and cannot move as they would in the real world; for example, they cannot move forward while facing right.
Furthermore, because the experimental environment repeats identical scenes, the participants could not visually identify their progress from the start location of each trial.
Thus, the navigation task in this experiment would have been more difficult than in a real-world environment.
\par
We have not tested the subjective workload because this paper aims to verify that haptic navigation and enhancing music-listening experience can be compatible.
However, nine participants (par 7, 8, 9, 12, 14, 18, 19, 20, 21) commented that direction finding with modulated musical vibrations required more concentration than voice localization, suggesting that the proposed method may have placed a high workload on some participants.
In addition, the improved music-listening experience can lead to distraction from one's surroundings.
The higher subjective workload of the proposed method will not be a major problem in VEs, but for real-world use it should be improved to reduce the subjective workload and notify users of dangers to prevent a serious accident.
\section{Conclusion}
This paper proposed a method to achieve both haptic navigation and enhance the music-listening experience by stimulating musical vibrations, modulated based on the positional relationship between the user and the target, on both sides of the neck using Hapbeat.
The results of the experiment confirmed that the stimulating musical vibrations enhanced the music-listening experience even during navigation, and although navigation by voice localization was easier, participants were able to reach the target using the proposed method.
The method can also convey distance information---which has been an issue with conventional navigation methods---by modulating the music without disturbing the listening experience.
In summary, while the proposed method is not suitable for all, it is a good option for the following users: those who dislike voice localization while listening to music, those whose tactile sense on the neck is sensitive, and those who wish to enjoy immersive music listening during navigation.
The results also showed that stimulating musical vibration by Hapbeat can enhance the listening experience without interfering with the conventional navigation methods, indicating that Hapbeat is useful in a broad variety of cases.
\par
The proposed method can be used for other applications such as video games, virtual reality, and video watching to present information---direct attention to specific objects in a complicated environment and show the presence of approaching humans---enhancing, or not disturbing, the experience.
Haptic technologies have often been proposed independently of enhancing the application experience and information presentation, but the proposed method can achieve both with an easy-to-use necklace-type device.
We hope the proposed method helps expand the use of haptic technologies and leads to their practical use in the future.

\section*{Acknowledgment}
This work was supported by JSPS KAKENHI Grant Numbers JP17H01774, JP20H04220.
\ifCLASSOPTIONcaptionsoff
  \newpage
\fi



\bibliography{bibmain}

\begin{thebibliography}{10}
\providecommand{\url}[1]{#1}
\csname url@samestyle\endcsname
\providecommand{\newblock}{\relax}
\providecommand{\bibinfo}[2]{#2}
\providecommand{\BIBentrySTDinterwordspacing}{\spaceskip=0pt\relax}
\providecommand{\BIBentryALTinterwordstretchfactor}{4}
\providecommand{\BIBentryALTinterwordspacing}{\spaceskip=\fontdimen2\font plus
\BIBentryALTinterwordstretchfactor\fontdimen3\font minus
  \fontdimen4\font\relax}
\providecommand{\BIBforeignlanguage}[2]{{%
\expandafter\ifx\csname l@#1\endcsname\relax
\typeout{** WARNING: IEEEtran.bst: No hyphenation pattern has been}%
\typeout{** loaded for the language `#1'. Using the pattern for}%
\typeout{** the default language instead.}%
\else
\language=\csname l@#1\endcsname
\fi
#2}}
\providecommand{\BIBdecl}{\relax}
\BIBdecl

\bibitem{haas2018can}
G.~Haas, E.~Stemasov, and E.~Rukzio, ``Can't you hear me? investigating
  personal soundscape curation,'' in \emph{Proceedings of the 17th
  International Conference on Mobile and Ubiquitous Multimedia}, 2018, pp.
  59--69.

\bibitem{jones2008ontrack}
M.~Jones, S.~Jones, G.~Bradley, N.~Warren, D.~Bainbridge, and G.~Holmes,
  ``Ontrack: Dynamically adapting music playback to support navigation,''
  \emph{Personal and Ubiquitous Computing}, vol.~12, no.~7, pp. 513--525, 2008.

\bibitem{yamano2012eyesound}
S.~Yamano, T.~Hamajo, S.~Takahashi, and K.~Higuchi, ``Eyesound: single-modal
  mobile navigation using directionally annotated music,'' in \emph{Proceedings
  of the 3rd Augmented Human International Conference}, 2012, pp. 1--4.

\bibitem{heller2018navigatone}
F.~Heller and J.~Sch{\"o}ning, ``Navigatone: Seamlessly embedding navigation
  cues in mobile music listening,'' in \emph{Proceedings of the 2018 CHI
  Conference on Human Factors in Computing Systems}, 2018, pp. 1--7.

\bibitem{heller2020attracktion}
F.~Heller, J.~Adamczyk, and K.~Luyten, ``Attracktion: Field evaluation of
  multi-track audio as unobtrusive cues for pedestrian navigation,'' in
  \emph{22nd International Conference on Human-Computer Interaction with Mobile
  Devices and Services}, 2020, pp. 1--7.

\bibitem{reybrouck2019music}
M.~Reybrouck, P.~Podlipniak, and D.~Welch, ``Music and noise: Same or
  different? what our body tells us,'' \emph{Frontiers in psychology}, vol.~10,
  p. 1153, 2019.

\bibitem{merchel2013music}
S.~Merchel and M.~E. Altinsoy, ``Music-induced vibrations in a concert hall and
  a church,'' \emph{Archives of Acoustics}, vol.~38, 2013.

\bibitem{takahashi2002some}
Y.~Takahashi, K.~Kanada, and Y.~Yonekawa, ``Some characteristics of human body
  surface vibration induced by low frequency noise,'' \emph{Journal of Low
  Frequency Noise, Vibration and Active Control}, vol.~21, no.~1, pp. 9--19,
  2002.

\bibitem{merchel2018auditory}
S.~Merchel and M.~E. Altinsoy, ``Auditory-tactile experience of music,'' in
  \emph{Musical Haptics}.\hskip 1em plus 0.5em minus 0.4em\relax Springer,
  Cham, 2018, pp. 123--148.

\bibitem{hove2020feel}
M.~J. Hove, S.~A. Martinez, and J.~Stupacher, ``Feel the bass: Music presented
  to tactile and auditory modalities increases aesthetic appreciation and body
  movement.'' \emph{Journal of Experimental Psychology: General}, vol. 149,
  no.~6, p. 1137, 2020.

\bibitem{yamazaki2016tension}
Y.~Yamazaki, H.~Mitake, and S.~Hasegawa, ``Tension-based wearable vibroacoustic
  device for music appreciation,'' in \emph{International conference on human
  haptic sensing and touch enabled computer applications}.\hskip 1em plus 0.5em
  minus 0.4em\relax Springer, 2016, pp. 273--283.

\bibitem{yamazaki2022implementation}
Y.~Yamazaki, H.~Mitake, and S.~Hasegawa, ``Implementation of tension-based
  compact necklace-type haptic device achieving widespread transmission of
  low-frequency vibrations,'' \emph{IEEE Transactions on Haptics}, 2022.

\bibitem{yamazaki2019neck}
Y.~Yamazaki, S.~Hasegawa, H.~Mitake, and A.~Shirai, ``Neck strap haptics: An
  algorithm for non-visible vr information using haptic perception on the
  neck,'' in \emph{ACM SIGGRAPH 2019 Posters}, 2019, pp. 1--2.

\bibitem{klatzky2006cognitive}
R.~L. Klatzky, J.~R. Marston, N.~A. Giudice, R.~G. Golledge, and J.~M. Loomis,
  ``Cognitive load of navigating without vision when guided by virtual sound
  versus spatial language.'' \emph{Journal of experimental psychology:
  Applied}, vol.~12, no.~4, p. 223, 2006.

\bibitem{holland2002audiogps}
S.~Holland, D.~R. Morse, and H.~Gedenryd, ``Audiogps: Spatial audio navigation
  with a minimal attention interface,'' \emph{Personal and Ubiquitous
  computing}, vol.~6, no.~4, pp. 253--259, 2002.

\bibitem{strachan2005gpstunes}
S.~Strachan, P.~Eslambolchilar, R.~Murray-Smith, S.~Hughes, and S.~O'Modhrain,
  ``Gpstunes: controlling navigation via audio feedback,'' in \emph{Proceedings
  of the 7th international conference on Human computer interaction with mobile
  devices \& services}, 2005, pp. 275--278.

\bibitem{kimura1964left}
D.~Kimura, ``Left-right differences in the perception of melodies,''
  \emph{Quarterly Journal of Experimental Psychology}, vol.~16, no.~4, pp.
  355--358, 1964.

\bibitem{bosman2003gentleguide}
S.~Bosman, B.~Groenendaal, J.-W. Findlater, T.~Visser, M.~de~Graaf, and
  P.~Markopoulos, ``Gentleguide: An exploration of haptic output for indoors
  pedestrian guidance,'' in \emph{International Conference on Mobile
  Human-Computer Interaction}.\hskip 1em plus 0.5em minus 0.4em\relax Springer,
  2003, pp. 358--362.

\bibitem{salzer2010vibrotactor}
Y.~Salzer, T.~Oron-Gilad, and A.~Ronen, ``Vibrotactor-belt on the
  thigh--directions in the vertical plane,'' in \emph{International Conference
  on Human Haptic Sensing and Touch Enabled Computer Applications}.\hskip 1em
  plus 0.5em minus 0.4em\relax Springer, 2010, pp. 359--364.

\bibitem{erp2005waypoint}
J.~B.~V. Erp, H.~A.~V. Veen, C.~Jansen, and T.~Dobbins, ``Waypoint navigation
  with a vibrotactile waist belt,'' \emph{ACM Transactions on Applied
  Perception (TAP)}, vol.~2, no.~2, pp. 106--117, 2005.

\bibitem{schaack2019haptic}
S.~Schaack, G.~Chernyshov, K.~Ragozin, B.~Tag, R.~Peiris, and K.~Kunze,
  ``Haptic collar: Vibrotactile feedback around the neck for guidance
  applications,'' in \emph{Proceedings of the 10th Augmented Human
  International Conference 2019}, 2019, pp. 1--4.

\bibitem{marston2007nonvisual}
J.~R. Marston, J.~M. Loomis, R.~L. Klatzky, and R.~G. Golledge, ``Nonvisual
  route following with guidance from a simple haptic or auditory display,''
  \emph{Journal of Visual Impairment \& Blindness}, vol. 101, no.~4, pp.
  203--211, 2007.

\bibitem{fujimoto2014non}
E.~Fujimoto and M.~Turk, ``Non-visual navigation using combined audio music and
  haptic cues,'' in \emph{Proceedings of the 16th International Conference on
  Multimodal Interaction}, 2014, pp. 411--418.

\bibitem{di2019haptic}
P.~Di~Campli San~Vito, G.~Shakeri, S.~Brewster, F.~Pollick, E.~Brown,
  L.~Skrypchuk, and A.~Mouzakitis, ``Haptic navigation cues on the steering
  wheel,'' in \emph{Proceedings of the 2019 CHI Conference on Human Factors in
  Computing Systems}, 2019, pp. 1--11.

\bibitem{chinello2017design}
F.~Chinello, C.~Pacchierotti, J.~Bimbo, N.~G. Tsagarakis, and D.~Prattichizzo,
  ``Design and evaluation of a wearable skin stretch device for haptic
  guidance,'' \emph{IEEE Robotics and Automation Letters}, vol.~3, no.~1, pp.
  524--531, 2017.

\bibitem{spiers2016design}
A.~J. Spiers and A.~M. Dollar, ``Design and evaluation of shape-changing haptic
  interfaces for pedestrian navigation assistance,'' \emph{IEEE transactions on
  haptics}, vol.~10, no.~1, pp. 17--28, 2016.

\bibitem{amemiya2009haptic}
T.~Amemiya and H.~Sugiyama, ``Haptic handheld wayfinder with pseudo-attraction
  force for pedestrians with visual impairments,'' in \emph{Proceedings of the
  11th international ACM SIGACCESS conference on Computers and accessibility},
  2009, pp. 107--114.

\bibitem{senn2020experience}
O.~Senn, T.~Bechtold, D.~Rose, G.~S. C{\^a}mara, N.~D{\"u}vel, R.~Jerjen,
  L.~Kilchenmann, F.~Hoesl, A.~Baldassarre, and E.~Alessandri, ``Experience of
  groove questionnaire: Instrument development and initial validation,''
  \emph{Music Perception: An Interdisciplinary Journal}, vol.~38, no.~1, pp.
  46--65, 2020.

\bibitem{series2011algorithms}
B.~Series, ``Algorithms to measure audio programme loudness and true-peak audio
  level,'' 2011.

\bibitem{heller2009multi}
F.~Heller, T.~Knott, M.~Weiss, and J.~Borchers, ``Multi-user interaction in
  virtual audio spaces,'' in \emph{CHI'09 Extended Abstracts on Human Factors
  in Computing Systems}, 2009, pp. 4489--4494.

\bibitem{gorzel2019efficient}
M.~Gorzel, A.~Allen, I.~Kelly, J.~Kammerl, A.~Gungormusler, H.~Yeh, and
  F.~Boland, ``Efficient encoding and decoding of binaural sound with resonance
  audio,'' in \emph{Audio Engineering Society Conference: 2019 AES
  International Conference on Immersive and Interactive Audio}.\hskip 1em plus
  0.5em minus 0.4em\relax Audio Engineering Society, 2019.

\bibitem{williamson2010social}
J.~Williamson, S.~Robinson, C.~Stewart, R.~Murray-Smith, M.~Jones, and
  S.~Brewster, ``Social gravity: a virtual elastic tether for casual,
  privacy-preserving pedestrian rendezvous,'' in \emph{Proceedings of the
  SIGCHI Conference on Human Factors in Computing Systems}, 2010, pp.
  1485--1494.

\bibitem{honda2016detection}
A.~Honda, K.~Ohba, Y.~Iwaya, and Y.~Suzuki, ``Detection of sound image movement
  during horizontal head rotation,'' \emph{i-Perception}, vol.~7, no.~5, p.
  2041669516669614, 2016.

\bibitem{holm1979simple}
S.~Holm, ``A simple sequentially rejective multiple test procedure,''
  \emph{Scandinavian journal of statistics}, pp. 65--70, 1979.

\bibitem{merchel2009hearing}
S.~Merchel, A.~Leppin, and E.~Altinsoy, ``Hearing with your body: the influence
  of whole-body vibrations on loudness perception,'' in \emph{Proceedings of
  the 16th International Congress on Sound and Vibration (ICSV), Krak{\'o}w,
  Poland}, vol.~4, 2009.

\bibitem{aevarsson2022vibrotactile}
E.~A. {\AE}varsson, T.~{\'A}sgeirsd{\'o}ttir, F.~Pind, {\'A}.~Kristj{\'a}nsson,
  and R.~Unnthorsson, ``Vibrotactile threshold measurements at the wrist using
  parallel vibration actuators,'' \emph{ACM Transactions on Applied Perceptions
  (TAP)}, 2022.

\end{thebibliography}
\bibliographystyle{IEEEtran}
%


%

%

\begin{IEEEbiography}[{\includegraphics[width=1in,height=1.25in,clip,keepaspectratio]{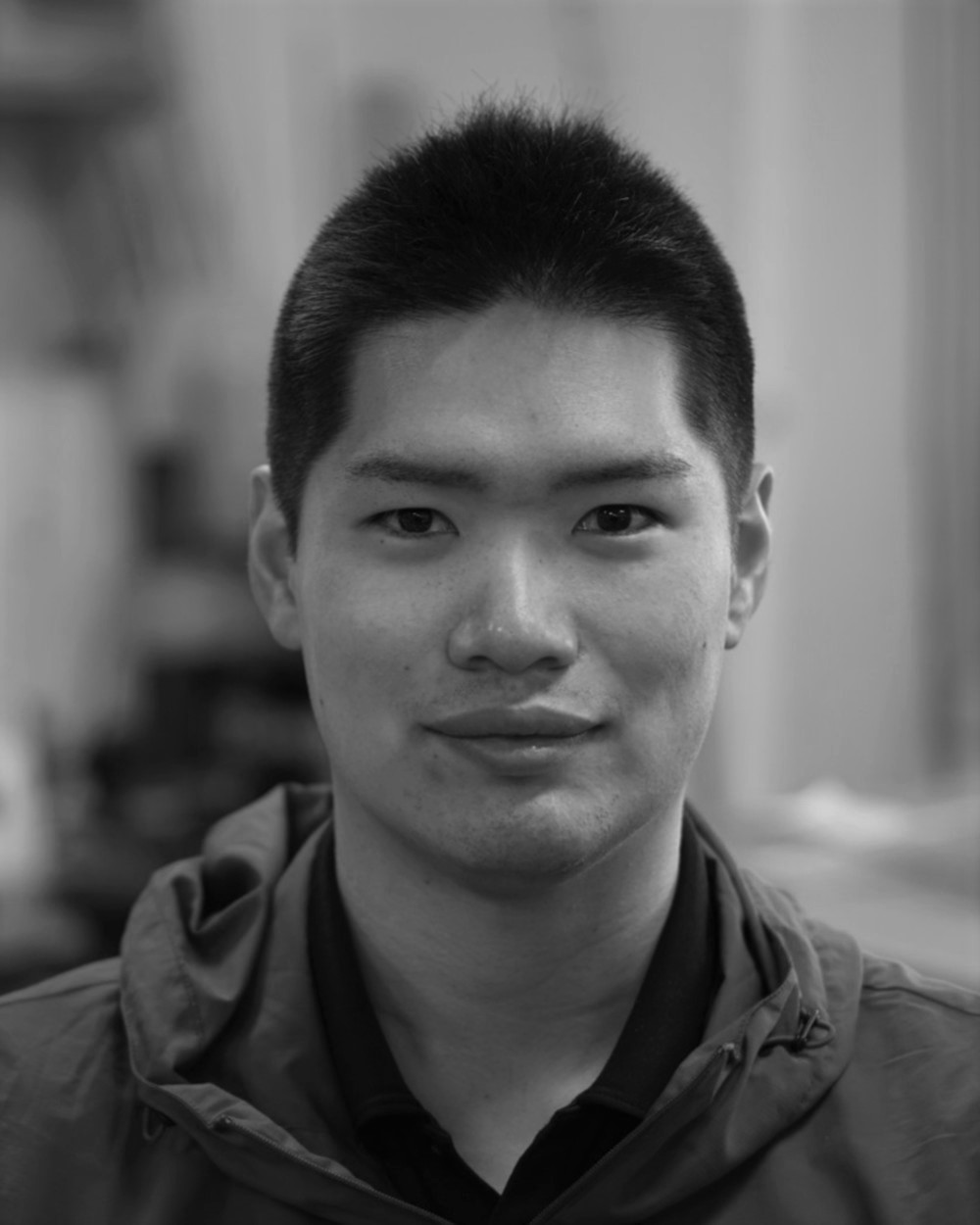}}]{Yusuke Yamazaki}
received the Ph.D. degree in information and communications engineering from Tokyo Institute of Technology in 2023. 
He established Hapbeat LLC. in 2017 to commercialize haptic device using invented vibration mechanism. 
His research interest is the popularization and social implementation of haptic technology relating to XR technologies and entertainment.
\end{IEEEbiography}

\begin{IEEEbiography}[{\includegraphics[width=1in,height=1.25in,clip,keepaspectratio]{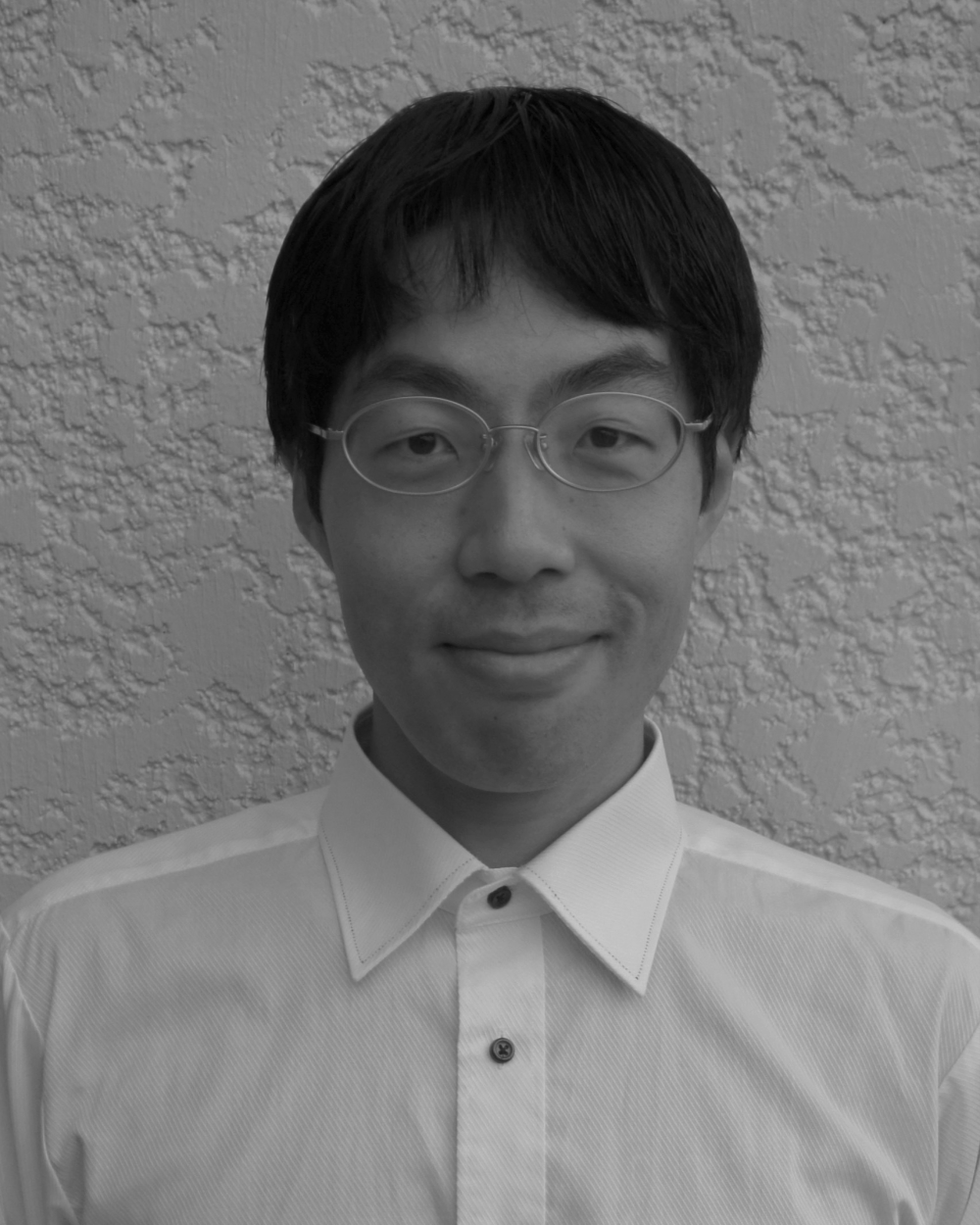}}]{Shoichi Hasegawa}
received the D.Eng. degree in computational intelligence and systems from the Tokyo Institute of Technology, Tokyo, Japan. He has been an Associate Professor with the Tokyo Institute of Technology since 2010 and was previously an Associate Professor with the University of Electro Communications. His domain of research includes haptic renderings, realtime simulations, interactive characters, soft and entertainment robotics, and virtual reality.
\end{IEEEbiography}


\vfill


\end{document}